\documentclass[aps, prd, twocolumn, showpacs, showkeys, superscriptaddress, preprintnumbers, floatfix]{revtex4}
\usepackage{multirow}
\usepackage{slashed}
\usepackage{graphicx}
\usepackage{amsmath}
\usepackage{bm}
\usepackage{yhmath}
\usepackage{hyperref}
\usepackage{subfigure}
\usepackage[T1]{fontenc}
\usepackage{color}
\usepackage{cases}
\usepackage{subfigure}
\usepackage{dcolumn, booktabs, bm}
\usepackage{amsfonts, amssymb, stmaryrd, latexsym, amsmath}
\usepackage{textcomp}
\usepackage{amsthm}
\usepackage{mathrsfs}

\usepackage{array}

\newcounter{RomanNumber}

\newcommand{\lyxmathsym}[1]{\ifmmode\begingroup\def\b@ld{bold}
	\text{\ifx\math@version\b@ld\bfseries\fi#1}\endgroup\else#1\fi}

\renewcommand{\arraystretch}{1.9}
\usepackage{braket}
\allowdisplaybreaks
\begin{document}
	\title{Axial charges and magnetic moments of the decuplet pentaquark family }
	
	\author{Hao-Song Li}\email{haosongli@nwu.edu.cn}\affiliation{School of Physics, Northwest University, Xi'an 710127, China}\affiliation{Shaanxi Key Laboratory for Theoretical Physics Frontiers, Xi'an 710127, China}\affiliation{Peng Huanwu Center for Fundamental Theory, Xi'an 710127, China}	
	\begin{abstract}
We present a systematic calculation of the axial charges and magnetic moments for the decuplet of hidden-charm molecular pentaquarks within the framework of the constituent quark model. Our findings reveal that the axial charges of these states are comparable in magnitude to that of the nucleon. Furthermore, we find that their magnetic moments obey a set of sum rules (denoted the Hao-Song sum rules) which hold for states sharing the same spin-flavor configuration, even in the presence of SU(3) flavor symmetry breaking. These results provide crucial insights into the internal structure and dynamics of multiquark hadrons, offering valuable guidance for future experimental and theoretical investigations.
\begin{eqnarray}
	&&	\mu_{P_{\psi}^{\Delta+}}-
	\mu_{P_{\psi}^{\Delta0}}+\mu_{P_{\psi s}^{\Sigma-}}-
	\mu_{P_{\psi s}^{\Sigma+}}
	+\mu_{P_{\psi ss}^{N0}}-
	\mu_{P_{\psi ss}^{N-}}=0, \nonumber\\
	&&	\mu_{P_{\psi}^{\Delta++}}+\mu_{P_{\psi}^{\Delta-}}
	+\mu_{P_{\psi sss}^{\Lambda-}}=
	3\mu_{P_{\psi s}^{\Sigma0}}. \nonumber
\end{eqnarray}

	\end{abstract}

	\maketitle
	\thispagestyle{empty}
	
	\section{Introduction}
	\label{sec1}
	
	The quark model, established by Gell-Mann in 1964 \cite{Gell-Mann:1964ewy}, provides a foundational framework for classifying hadrons into singlets, octets, and decuplets under SU(3) flavor symmetry. Beyond successfully describing conventional mesons and baryons, the model naturally accommodates the existence of exotic multiquark states. Decades of experimental progress have now confirmed this prediction, with numerous tetraquark and pentaquark candidates observed in various experiments \cite{LHCb:2016axx,BESIII:2016adj,LHCb:2018oeg,LHCb:2015yax,LHCb:2019kea,LHCb:2021chn,LHCb:2020jpq,LHCb:2022ogu,Belle:2025rso}. In this paper, we adopt the updated nomenclature for these pentaquark states as detailed in Ref. \cite{Gershon:2022xnn}.
	
	A landmark achievement in this field was the 2015 discovery by the LHCb collaboration of the first hidden-charm pentaquark candidates, $P^{N}_{\psi}(4380)^{^{+}}$ and $P^{N}_{\psi}(4450)^{^{+}}$, in the decay $\Lambda^{0}_b\to{}J/\psi{}K^{-}p$ ~\cite{LHCb:2015yax}. In 2019, an updated analysis by LHCb revealed three additional states: $P^{N}_{\psi}(4312)^{^{+}}$, $P^{N}_{\psi}(4440)^{^{+}}$  and $P^{N}_{\psi}(4457)^{^{+}}$~\cite{LHCb:2019kea}. Notably, the previously observed  $P^{N}_{\psi}(4450)^{^{+}}  $ was resolved into two distinct peaks corresponding to $P^{N}_{\psi}(4440)^{^{+}}  $  and the $P^{N}_{\psi}(4457)^{^{+}}$. Further expanding the family, LHCb observed the hidden-charm strange pentaquark $P_{\psi{}s}^{\Lambda}(4459)^{0}$ in 2020 via amplitude analysis of $\Xi^{-}_b\to{}J/\psi{}\Lambda{}K^{-}$ \cite{LHCb:2020jpq}. In 2021, a new pentaquark structure, $P^{N}_{\psi}(4337)^{^{+}}  $, was identified in$B_{s}^{0}\to J/\psi p\bar{p}$decays~\cite{LHCb:2021chn}. Most recently, LHCb and Belle reported the observation of $P_{\psi{}s}^{\Lambda}(4338)^{0}$ in $B^{-}\to{}J/\psi{}\Lambda\bar{p}$ decays and $\Upsilon(1S, 2S)$ inclusive decays to $J/\psi\Lambda$~\cite{LHCb:2022ogu,Belle:2025rso}.

	The experimental discoveries of hidden-charm pentaquarks have stimulated extensive theoretical work to determine their spin-parity quantum numbers ~\cite{Deng:2022vkv,Wang:2019nvm,Xiao:2019gjd,Liu:2020hcv,Peng:2020hql,Zhu:2021lhd,Du:2021bgb,Chen:2021tip,Yang:2021pio,Guo:2017jvc,Dong:2021juy,Yang:2024okq,Roca:2025zyi,Liu:2025slt,Clymton:2025zer}, although definitive experimental assignments remain unavailable. A striking commonality among these states is their mass proximity to thresholds for conventional baryon-meson pair production. This near-threshold behavior strongly suggests a molecular structure, where the pentaquark is a bound state of a charmed baryon and an anti-charmed meson, potentially fitting into the octet or decuplet representations of the quark model. While this molecular picture provides a compelling theoretical basis, elucidating the exact nature of these states requires a deeper investigation of their fundamental properties.
	
	Among the most revealing hadronic properties are the axial charge and the magnetic moment. The axial charge, $g_A$, is a fundamental parameter of the Standard Model, bridging electroweak and strong interactions \cite{UCNA:2012fhw,Mund:2012fq}. It governs weak interaction processes such as beta decay, while also revealing deep connections between weak and strong interactions. This interplay is clearly embodied in the Goldberger-Treiman relation \cite{Goldberger:1958tr}, $g_A=f_{\pi}g_{\pi NN}/M_N$, which directly links the axial charge to the pion decay constant $f_{\pi}$, the pion-nucleon coupling $g_{\pi NN}$, and the nucleon mass $M_N$. As a result, the relationship between pion dynamics and axial charges lies at the heart of hadronic structure studies. Moreover, $g_A$ serves as a clear indicator of chiral symmetry breaking in non-perturbative QCD, making it an essential parameter in low-energy effective theories. Theoretical and phenomenological studies of axial charges have been actively pursued over the years~\cite{Bijnens:1984ec,Jenkins:1991es,Zhu:2002tn,Choi:2010ty,Dahiya:2023izc}. Therefore, investigating the axial charges of the hidden-charm pentaquark offers a complementary and independent approach to understanding these exotic states.
	
Beyond what can be learned from axial charge alone, exploring the properties of the hidden-charm pentaquark family within molecular models offers valuable insights into the internal structure of these exotic states. Electromagnetic properties of hadrons play a crucial role in deciphering strong interaction mechanisms and structural features, thereby shedding light on non-perturbative aspects of QCD in the low-energy regime. Among these properties, the magnetic moment provides key information on internal charge and magnetization distributions. The magnetic moments of hidden-charm pentaquark states have been investigated using quark-based approaches in Refs.~\cite{Gao:2021hmv,Wang:2016dzu,Guo:2023fih}. With the framework of QCD light-cone sum rules, the authors extracted the magnetic dipole moment of the	$P^{N}_{\psi}(4312)^{^{+}}$,	$P^{N}_{\psi}(4440)^{^{+}}$, $P^{N}_{\psi}(4457)^{^{+}}$, $P^{N}_{\psi}(4380)^{^{+}}$ and 	$P_{\psi{}s}^{\Lambda}(4459)^{0}$ pentaquark states in the molecular and diquark-diquark-antiquark models in Ref.~\cite{Ozdem:2021ugy}. In Ref.~\cite{Ortiz-Pacheco:2018ccl}, the authors determined the ground state configuration of hidden-charm pentaquark states with quantum numbers $J^P = \frac{3}{2}^-$, along with the corresponding magnetic dipole moments and electromagnetic coupling constants. These results provide valuable input for quark-model-based studies of pentaquark photoproduction processes. 
	
	The investigation of axial charges and magnetic moments of pentaquark states offers crucial insight into their internal structure and the interquark dynamics governed by the strong interaction. These observables shed light on how quarks are bound into composite hadrons through non-perturbative QCD mechanisms. With ongoing and future experimental advancements, the discovery of the complete octet and decuplet of hidden-charm molecular pentaquarks appears increasingly attainable. Such progress will intensify interest in deciphering the internal organization of these states, where studies of axial charges and magnetic moments can provide essential quantitative guidance.

	In this work, we extend our previous analysis of the octet hidden-charm molecular pentaquarks \cite{Li:2024wxr, Lei:2024geu} to the decuplet representation. We report a comprehensive quark-model calculation of the axial charges and magnetic moments for the complete decuplet family. Our results demonstrate that the axial charges are of the same order as the nucleon's, and that the magnetic moments follow characteristic patterns, culminating in the Hao-Song sum rules that persist even under SU(3) symmetry breaking.
	
This paper is structured as follows.  Section \ref{sec2} introduces the molecular model and the corresponding wave functions. The axial charges of the decuplet hidden-charm molecular pentaquark are then calculated in Section \ref{sec3}, followed by the computation of their magnetic moments in Section \ref{sec4}. Finally, Section \ref{sec5} provides a summary of our work.

	\section{Wave Functions}
	\label{sec2}
The wave function $\Psi$ of a hadronic state is central to the computation of its properties. In the quark model, the total wave function is constructed as a direct product of its constituent parts: the flavor wave function $\psi_{flavor}  $, spin wave function $ \chi_{spin} $,  color wave function $ \xi_{color} $ and space wave function $ \eta_{space} $, each contributing to the complete quantum mechanical description. This is represented as:
	\begin{align}
	\Psi=\psi_{flavor}\otimes\chi_{spin}\otimes\xi_{color}\otimes\eta_{space}.
	\end{align}
	
The overall wave function must be antisymmetric under fermion exchange, as required by Pauli exclusion principle. In the ground state, where the color wave function $\xi_{\text{color}}$ is antisymmetric and the spatial wave function $\eta_{\text{space}}$ is symmetric, the flavor-spin component must be constructed to satisfy the overall antisymmetry condition in calculations of axial charges and magnetic moments.

In the molecular picture, the hidden-charm pentaquark family is composed of a singly charmed baryon and an anti-charmed meson. The singly charmed baryons can be derived from light baryons under $SU(3)$ flavor symmetry by replacing one light quark with a charm quark. Depending on the symmetry of the two remaining light quarks, two configurations arise:
\begin{itemize}
	\item When the light quark pair is symmetric, the charmed baryon belongs to the $\mathbf{6}_f$ representation. Combining this with the anti-charmed meson ($\overline{\mathbf{3}}_f$) yields pentaquark states in the $\mathbf{10}_f$ and $\mathbf{8}_{1f}$ representations.
	
	\item When the light quark pair is antisymmetric, the charmed baryon lies in the $\overline{\mathbf{3}}_f$ representation, leading to pentaquarks in the $\mathbf{8}_{2f}$ and $\mathbf{1}_f$ representations when coupled with the anti-charmed meson.
\end{itemize}

		\renewcommand{\arraystretch}{1.6}
	\begin{table}
		\centering
		\caption{The flavor wave functions $\psi_{flavor}$ and the spin wave functions $\chi_{spin}$ of the $S$-wave anti-charmed mesons. Here, $S$ and $S_3$ are the spin and its third component, while the arrow denotes the third component of the quark spin.}
		\label{tab_1}
		\begin{tabular}{c|l|l}
			\toprule[1.0pt]
			\toprule[1.0pt]
			Mesons
			&	$\ket{S,S_3}$
			&	$\psi_{flavor} \otimes \chi_{spin}$
			\\
			\hline
			
			$\bar{D}^{0}$
			&	$\ket{0,0}$
			&	$\bar{c} u \otimes {\frac{1}{\sqrt{2}}}(\uparrow\downarrow-\downarrow\uparrow)$
			\\
			\hline
			$D^{-}$
			&	$\ket{0,0}$
			&	$\bar{c} d \otimes {\frac{1}{\sqrt{2}}}(\uparrow\downarrow-\downarrow\uparrow)$
			
			\\
			\hline
			$D_{s}^{-}$
			&	$\ket{0,0}$
			&	$\bar{c} s \otimes {\frac{1}{\sqrt{2}}}(\uparrow\downarrow-\downarrow\uparrow)$
			\\
			\hline
			
			\multirow{3}{*}{$\bar{D}^{*0}$}
			&	$\ket{1,1}$
			&	$\bar{c} u \otimes \uparrow\uparrow$
			\\

			&	$\ket{1,0}$
			&	$\bar{c} u \otimes {\frac{1}{\sqrt{2}}}(\uparrow\downarrow + \downarrow\uparrow)$
			\\
			
			&	$\ket{1,-1}$
			&	$\bar{c} u \otimes \downarrow\downarrow$
			\\
			\hline
			
			\multirow{3}{*}{$D^{*-}$}
			&	$\ket{1,1}$
			&	$\bar{c} d \otimes \uparrow\uparrow$
			\\

			&	$\ket{1,0}$
			&	$\bar{c} d \otimes {\frac{1}{\sqrt{2}}}(\uparrow\downarrow + \downarrow\uparrow)$
			\\
			
			&	$\ket{1,-1}$
			&	$\bar{c} d \otimes \downarrow\downarrow$
			\\
			\hline
			
			\multirow{3}{*}{$D_{s}^{*-}$}
			&	$\ket{1,1}$
			&	$\bar{c} s \otimes \uparrow\uparrow$
			\\

			&	$\ket{1,0}$
			&	$\bar{c} s \otimes {\frac{1}{\sqrt{2}}}(\uparrow\downarrow + \downarrow\uparrow)$
			\\

			&	$\ket{1,-1}$
			&	$\bar{c} s \otimes \downarrow\downarrow$
			\\
			\toprule[1.0pt]
			\toprule[1.0pt]
		\end{tabular}
	\end{table}		
	This structure follows from the group decomposition $ \mathbf{3} \otimes \mathbf{3} \otimes \mathbf{3} = \mathbf{1} \oplus \mathbf{8}_1 \oplus \mathbf{8}_2 \oplus \mathbf{10} $. We extend our study of hidden-charm molecular pentaquarks from the octet \cite{Li:2024wxr} to the decuplet by performing a systematic quark-model calculation of their axial charges and magnetic moments. The decuplet hidden-charm molecular pentaquark states have three spin configurations $J^{P}(J_{b} ^{P_{b}}\otimes J_{m} ^{P_{m}})  $: $\frac{1}{2}^{-}(\frac{1}{2}^{+}\otimes0^{-})  $, $\frac{1}{2}^{-}(\frac{1}{2}^{+}\otimes1^{-})  $, and $\frac{3}{2}^{-}(\frac{1}{2}^{+}\otimes1^{-})  $,  where $ J^{P} $ is the total spin of  pentaquark states, and   $J_{b} ^{P_{b}}\otimes J_{m} ^{P_{m}} $ correspond to the angular momentum and parity of baryon and meson, respectively. In Table \ref{tab_1}, we present the flavor wave functions $\psi_{flavor}$ and the spin wave functions $\chi_{spin}$ of the $ S $-wave anti-charmed mesons. In Table \ref{tab_2}, we present the flavor wave functions $\psi_{flavor}$ and the spin wave functions $\chi_{spin}$ of the $S$-wave singly charmed baryons. The flavor wave functions of the hidden-charm molecular pentaquark under $SU(3)$ symmetry are listed in Table \ref{tab_3}.
	
		\renewcommand{\arraystretch}{1.65}
		\begin{table}[htp]
			\centering
			\caption{The flavor wave functions $\psi_{flavor}$ and  the spin wave functions $\chi_{spin}$ of the $S$-wave singly charmed baryon. Here, $S$ and $S_3$ are the spin and its third component, while	the arrow denotes the third component of the quark spin.}
			\label{tab_2}
			\begin{tabular}{c|l|l}
			\toprule[1.0pt]
			\toprule[1.0pt]
				Baryons
				&	$\ket{S,S_3}$
				&	$\psi_{flavor} \otimes \chi_{spin}$
				\\
				\hline
		
				\multirow{2}{*}{$\Sigma_{c}^{++}$}
				&	$\ket{\frac{1}{2},\frac{1}{2}}$
				&	$uuc \otimes {\frac{1}{\sqrt{6}}}(2\uparrow\uparrow\downarrow-\downarrow\uparrow\uparrow-\uparrow\downarrow\uparrow)$
				\\
				
				&	$\ket{\frac{1}{2},-\frac{1}{2}}$
				&   $uuc \otimes {\frac{1}{\sqrt{6}}}(\uparrow\downarrow\downarrow+\downarrow\uparrow\downarrow-2\downarrow\downarrow\uparrow)$
				\\
				\hline
		
				\multirow{2}{*}{$\Sigma_{c}^{+}$ }
				&	$\ket{\frac{1}{2},\frac{1}{2}}$
				&   ${\frac{1}{\sqrt{2}}}(udc + duc) \otimes {\frac{1}{\sqrt{6}}}(2\uparrow\uparrow\downarrow-\downarrow\uparrow\uparrow-\uparrow\downarrow\uparrow)$
				\\

				&	$\ket{\frac{1}{2},-\frac{1}{2}}$
				&   ${\frac{1}{\sqrt{2}}}(udc + duc) \otimes {\frac{1}{\sqrt{6}}}(\uparrow\downarrow\downarrow+\downarrow\uparrow\downarrow-2\downarrow\downarrow\uparrow)$
				\\
				\hline
		
				\multirow{2}{*}{$\Sigma_{c}^{0}$}
				&	$\ket{\frac{1}{2},\frac{1}{2}}$
				&	$ddc \otimes {\frac{1}{\sqrt{6}}}(2\uparrow\uparrow\downarrow-\downarrow\uparrow\uparrow-\uparrow\downarrow\uparrow)$
				\\

				&	$\ket{\frac{1}{2},-\frac{1}{2}}$
				&   $ddc \otimes {\frac{1}{\sqrt{6}}}(\uparrow\downarrow\downarrow+\downarrow\uparrow\downarrow-2\downarrow\downarrow\uparrow)$
				\\
				\hline
		
				\multirow{2}{*}{$\Xi_{c}^{\prime +}$ }
				&	$\ket{\frac{1}{2},\frac{1}{2}}$
				&   ${\frac{1}{\sqrt{2}}}(usc + suc) \otimes {\frac{1}{\sqrt{6}}}(2\uparrow\uparrow\downarrow-\downarrow\uparrow\uparrow-\uparrow\downarrow\uparrow)$
		     	\\

				&	$\ket{\frac{1}{2},-\frac{1}{2}}$
				&   ${\frac{1}{\sqrt{2}}}(usc + suc) \otimes {\frac{1}{\sqrt{6}}}(\uparrow\downarrow\downarrow+\downarrow\uparrow\downarrow-2\downarrow\downarrow\uparrow)$
				\\
				\hline
		
				\multirow{2}{*}{$\Xi_{c}^{\prime 0}$ }
				&	$\ket{\frac{1}{2},\frac{1}{2}}$
				&   ${\frac{1}{\sqrt{2}}}(dsc + sdc) \otimes {\frac{1}{\sqrt{6}}}(2\uparrow\uparrow\downarrow-\downarrow\uparrow\uparrow-\uparrow\downarrow\uparrow)$
				\\
		
				&	$\ket{\frac{1}{2},-\frac{1}{2}}$
				&   ${\frac{1}{\sqrt{2}}}(dsc + sdc) \otimes {\frac{1}{\sqrt{6}}}(\uparrow\downarrow\downarrow+\downarrow\uparrow\downarrow-2\downarrow\downarrow\uparrow)$
				\\
				\hline
				
				\multirow{2}{*}{$\Omega_{c}^{0}$}
				&	$\ket{\frac{1}{2},\frac{1}{2}}$
				&	$ssc \otimes {\frac{1}{\sqrt{6}}}(2\uparrow\uparrow\downarrow-\downarrow\uparrow\uparrow-\uparrow\downarrow\uparrow)$
				\\
		
				&	$\ket{\frac{1}{2},-\frac{1}{2}}$
				&   $ssc \otimes {\frac{1}{\sqrt{6}}}(\uparrow\downarrow\downarrow+\downarrow\uparrow\downarrow-2\downarrow\downarrow\uparrow)$
				\\
				\hline
		
				\multirow{2}{*}{$\Xi_{c}^{+}$ }
				&	$\ket{\frac{1}{2},\frac{1}{2}}$
				&   ${\frac{1}{\sqrt{2}}}(usc - suc) \otimes {\frac{1}{\sqrt{2}}}(\uparrow\downarrow\uparrow-\downarrow\uparrow\uparrow)$
				\\
		
				&	$\ket{\frac{1}{2},-\frac{1}{2}}$
				&   ${\frac{1}{\sqrt{2}}}(usc - suc) \otimes {\frac{1}{\sqrt{2}}}(\uparrow\downarrow\downarrow-\downarrow\uparrow\downarrow)$
				\\
				\hline
		
				\multirow{2}{*}{$\Xi_{c}^{0}$ }
				&	$\ket{\frac{1}{2},\frac{1}{2}}$
				&   ${\frac{1}{\sqrt{2}}}(dsc - sdc) \otimes {\frac{1}{\sqrt{2}}}(\uparrow\downarrow\uparrow-\downarrow\uparrow\uparrow)$
				\\
		
				&	$\ket{\frac{1}{2},-\frac{1}{2}}$
				&   ${\frac{1}{\sqrt{2}}}(dsc - sdc) \otimes {\frac{1}{\sqrt{2}}}(\uparrow\downarrow\downarrow-\downarrow\uparrow\downarrow)$
				\\
				\hline
		
				\multirow{2}{*}{$\Lambda_{c}^{+}$ }
				&	$\ket{\frac{1}{2},\frac{1}{2}}$
				&   ${\frac{1}{\sqrt{2}}}(udc - duc) \otimes {\frac{1}{\sqrt{2}}}(\uparrow\downarrow\uparrow-\downarrow\uparrow\uparrow)$
				\\
		
				&	$\ket{\frac{1}{2},-\frac{1}{2}}$
				&   ${\frac{1}{\sqrt{2}}}(udc - duc) \otimes {\frac{1}{\sqrt{2}}}(\uparrow\downarrow\downarrow-\downarrow\uparrow\downarrow)$
				\\
			\toprule[1.0pt]
			\toprule[1.0pt]
			\end{tabular}
		\end{table}

		\renewcommand{\arraystretch}{1.65}
		\begin{table}[htbp]
			\centering
			\caption{The expressions for different flavor wave function of the decuplet hidden-charm molecular pentaquark states under $SU(3)$ symmetry in the molecular model.}
			\label{tab_3} 		
			\begin{tabular}{c|c|c}
				\toprule[1.0pt]
				\toprule[1.0pt]
			
			$(Y, I, I_3)$ & $10_f$ &Wave function
			\\
			\midrule[1pt] $(1,\frac{3}{2},\frac{3}{2})$  &  ${P_{\psi}^{\Delta++}}$& ${\bar D^{(*)0}}\Sigma_c^{++}$
			\\
			$(1,\frac{3}{2},\frac{1}{2})$  &  ${P_{\psi}^{\Delta+}}$  &  $\sqrt{\frac{2}{3}}{\bar D^{(*)0}}\Sigma_c^{+}+\sqrt{\frac{1}{3}}{D^{(*)-}}\Sigma_c^{++}$  \\
			$(1,\frac{3}{2},-\frac{1}{2})$  &  ${P_{\psi}^{\Delta0}}$&  $\sqrt{\frac{2}{3}}{ D^{(*)-}}\Sigma_c^{+}+\sqrt{\frac{1}{3}}{\bar D^{(*)0}}\Sigma_c^{0}$  \\
			$(1,\frac{3}{2},-\frac{3}{2})$  &  ${P_{\psi}^{\Delta-}}$&  ${ D^{(*)-}}\Sigma_c^{0}$ \\
			$ (0,1,1)$  &  ${P_{\psi s}^{\Sigma+}}$& $\sqrt{\frac{2}{3}}{\bar D^{(*)0}}\Xi_c^{\prime +}+\sqrt{\frac{1}{3}}{ D_s^{(*)-}}\Sigma_c^{++}$       \\
			$ (0,1,0)$  &  ${P_{\psi s}^{\Sigma0}}$& $\sqrt{\frac{1}{3}}({ D^{(*)-}}\Xi_c^{\prime +}+
			{\bar D^{(*)0}}\Xi_c^{\prime 0})+\sqrt{\frac{1}{3}}{D_s^{(*)-}}\Sigma_c^{+}$        \\
			$ (0,1,-1)$  &  ${P_{\psi s}^{\Sigma-}}$& $\sqrt{\frac{2}{3}}{ D^{(*)-}}\Xi_c^{\prime 0}+\sqrt{\frac{1}{3}}{ D_s^{(*)-}}\Sigma_c^{0}$        \\
			$ (-1,\frac{1}{2},\frac{1}{2})$  &  ${P_{\psi ss}^{N0}}$& $\sqrt{\frac{1}{3}}{\bar D^{(*)0}}\Omega_c^{0}+\sqrt{\frac{2}{3}}{D_s^{(*)-}}\Xi_c^{\prime +}$        \\
			$ (-1,\frac{1}{2},-\frac{1}{2})$  &  ${P_{\psi ss}^{N-}}$& $\sqrt{\frac{1}{3}}{D^{(*)-}}\Omega_c^{0}+\sqrt{\frac{2}{3}}{ D_s^{(*)-}}\Xi_c^{\prime 0}$          \\
			$(-2,0,0)$  &  ${P_{\psi sss}^{\Lambda -}}$&  ${D_s^{(*)-}}\Omega_c^{0}$  \\
			\toprule[1.0pt]
			\toprule[1.0pt]
		\end{tabular}
		\end{table}

	\section{Axial Charge of the decuplet Hidden-Charm Molecular Pentaquark Family }
	\label{sec3}

	The axial charge represents a fundamental quantity in the standard model, bridging electroweak and strong interaction physics. It also provides a key probe of spontaneous chiral symmetry breaking in the non-perturbative regime of QCD. Investigating the axial charges of the newly observed hidden-charm pentaquark states thus offers a promising avenue to shed new light on chiral symmetry realization in exotic hadronic systems.
	
At the quark level, the pion-quark interaction reads
		\begin{equation}
			\mathscr{L}_{quark}
			=\frac{1}{2} g_{q} \bar \psi_q\gamma^{\mu}\gamma_5
			\partial_{\mu}\phi \psi_q,
		\end{equation}
	where $g_q$ is the coupling constant at the quark level, $\sigma_z$ is the z-component Pauli matrix. $f_\pi=92$ MeV is the decay
	constant of the pseudoscalar meson. $\phi$ is the pseudoscalar meson field. Considering only the z-component, the $\pi_{0}$ quark interaction reads
		\begin{equation}
			\mathscr{L}_{quark}= \frac{1}{2}\frac{g_q}{f_\pi} (\bar u
			\sigma_z\partial_z\pi_0 u-\bar d \sigma_z\partial_z\pi_0 d),
		\end{equation}
	The nucleon-pion Lagrangian at hadron level reads
		\begin{equation}
			\mathscr{L}_{N}=\frac{1}{2}g_{A} \bar N \gamma^{\mu} \gamma_5 \partial_{\mu}\phi N,
		\end{equation}	
		where $g_{A}$ is the axial charge of the nucleon. Considering only the z-component, the nucleon-pion Lagrangian reads
		\begin{equation}
			\mathscr{L}_{N}= \frac{g_{A}}{f_\pi}\bar N\frac{\Sigma_{z}}{2}\partial_z \pi_0 N,
		\end{equation}
thus,
		\begin{equation}
			\langle  N, j_3=\frac{1}{2}; ~\pi_0|\mathscr{L}_{N}|N, j_3=\frac{1}{2}\rangle = \frac{1}{2}\frac{q_z}{f_\pi}g_{A},\label{equ:10}
		\end{equation}
		where $q_z$ denotes the z-component of the external momentum carried by $\pi_0$. 
		At the quark level,
		\begin{equation}
			\langle  N , j_3=\frac{1}{2} ;~\pi_0|	\mathscr{L}_{quark}|N, j_3=\frac{1}{2}\rangle=\frac{5}{6}  \frac{q_z}{f_\pi}g_q.\label{equ:11}
		\end{equation}
	 Within the framework of quark-hadron duality, we obtain the relation $g_q = \frac{3}{5}g_{A}$ from Eq. (\ref{equ:10}) and Eq. (\ref{equ:11}).

    The decuplet pentaquark Lagrangian with  $J^{P}=\frac{1}{2}^{-}(\frac{1}{2}^{+}\otimes0^{-})$ in $\mathbf{10}_f$ flavor representation reads
		\begin{eqnarray}
			\mathscr{L}^{\frac{1}{2}}_{1}
			&=\rm Tr\Big(d_{1}\bar{\mathcal{P}} \gamma_{\mu}\gamma^{5}\partial^{\mu}\Phi \mathcal{P}\Big),\label{equ:13}	
		\end{eqnarray}
	where $d_{1}$ is the axial coupling constant of pentaquark state with  $J^{P}=\frac{1}{2}^{-}(\frac{1}{2}^{+}\otimes0^{-})$ in $\mathbf{10}_f$ flavor representation. $\Phi$ represents the pseudoscalar meson field in $SU(3)$ flavor symmetry
		\begin{align}
			\Phi \equiv
			\left(
			\begin{array}{ccc}
				\pi_{0} + \frac{1}{\sqrt{3}}\eta
				&\sqrt{2}\pi^{+}
				&\sqrt{2}K^{+}
				\\
				\sqrt{2}\pi^{-}
				&-\pi_0 + \frac{1}{\sqrt{3}}\eta
				&\sqrt{2}K^{0}
				\\	
				\sqrt{2}K^{-}
				&\sqrt{2}\bar{K}^{0}
				&-\frac{2}{\sqrt{3}}\eta
			\end{array}
			\right).
		\end{align}
$\mathcal{P}$ represents the decuplet hidden-charm molecular pentaquark states. We adopt the tensor field $\mathcal{P}\equiv
	\mathcal{P}^{abc}$:
	\begin{eqnarray}
		&&\mathcal{P}^{111}=P_{\psi}^{\Delta++},\mathcal{P}^{112}=\frac{1}{\sqrt{3}}P_{\psi}^{\Delta+},
		\mathcal{P}^{122}=\frac{1}{\sqrt{3}}P_{\psi}^{\Delta0},\nonumber\\
		&&\mathcal{P}^{222}=P_{\psi}^{\Delta-},
		\mathcal{P}^{113}=\frac{1}{\sqrt{3}}P_{\psi s}^{\Sigma+},
		\mathcal{P}^{123}=\frac{1}{\sqrt{6}}P_{\psi s}^{\Sigma0},	\nonumber\\
		&&\mathcal{P}^{223}=\frac{1}{\sqrt{3}}P_{\psi s}^{\Sigma-},
		\mathcal{P}^{133}=\frac{1}{\sqrt{3}}P_{\psi ss}^{N0},\mathcal{P}^{233}=\frac{1}{\sqrt{3}}P_{\psi ss}^{N-},\nonumber\\
		&&	\mathcal{P}^{333}=P_{\psi sss}^{\Lambda -}.\label{equ:20}
	\end{eqnarray}

	Substituting the matrix representations of $\Phi$ and $\mathcal{P}$ into Eq. (\ref{equ:13}) and retaining only the $\pi_{0}$ meson term, we obtain
	\begin{align}
		\mathscr{L}^{\frac{1}{2}}_{10}
		&=
	{\frac{1}{f_{\pi}}} d_{1} {\bar{P}_{\psi}^{\Delta++}}\Sigma_{z}\partial_{z}{\pi^0}{P_{\psi}^{\Delta++}}+	{\frac{1}{3f_{\pi}}} d_{1} {\bar{P}_{\psi}^{\Delta+}}\Sigma_{z}\partial_{z}{\pi^0}{P_{\psi}^{\Delta+}}
		\nonumber
		\\
		&-{\frac{1}{3f_{\pi}}} d_{1} {\bar{P}_{\psi}^{\Delta0}}\Sigma_{z}\partial_{z}{\pi^0}{P_{\psi}^{\Delta0}}-	{\frac{1}{f_{\pi}}} d_{1} {\bar{P}_{\psi}^{\Delta-}}\Sigma_{z}\partial_{z}{\pi^0}{P_{\psi}^{\Delta-}}
		\nonumber\\
		&+{\frac{2}{3f_{\pi}}} d_{1} {\bar{P}_{\psi s}^{\Sigma+}}\Sigma_{z}\partial_{z}{\pi^0}{P_{\psi s}^{\Sigma+}}
		-	{\frac{2}{3f_{\pi}}} d_{1} {\bar{P}_{\psi s}^{\Sigma-}}\Sigma_{z}\partial_{z}{\pi^0}{P_{\psi s}^{\Sigma-}}
		\nonumber\\
		&+
			{\frac{1}{3f_{\pi}}} d_{1} {\bar{P}_{\psi ss}^{N0}}\Sigma_{z}\partial_{z}{\pi^0}{P_{\psi ss}^{N0}}-	{\frac{1}{3f_{\pi}}} d_{1} {\bar{P}_{\psi ss}^{N-}}\Sigma_{z}\partial_{z}{\pi^0}{P_{\psi ss}^{N-}}.\label{equ:21}
	\end{align}
		Having established the axial coupling constants for the decuplet hidden-charm molecular pentaquarks in terms of the parameter $d_{1}$, we note that Eq. (\ref{equ:21}) specifically details the components for the $\pi_{0}$ meson decay. The same formalism can be extended to derive the corresponding couplings for decays involving other light mesons, all of which are likewise determined by the same universal parameter $d_{1}$.

		 To determine the coupling constant $d_{1}$, similar to the procedure employed for the nucleon, we consider the $\pi_{0}$ meson decay of $P_{\psi}^{\Delta++}$ in Eq. (\ref{equ:21}). The Lagrangian for the $\pi_{0}$ decay of $P_{\psi}^{\Delta++}$ with the spin configuration $J^{P}=\frac{1}{2}^{-}(\frac{1}{2}^{+}\otimes0^{-})$ at the hadron level reads
		\begin{align}
			\mathscr{L}^{\frac{1}{2}}_{P_{\psi}^{\Delta++}}
			&=
			\frac{d_{1}}{f_\pi}{\bar{P}_{\psi}^{\Delta++}}\frac{\Sigma_z}{2}\partial_{z}  {\pi_{0}}{P_{\psi}^{\Delta++}}.
		\end{align}
		$\frac{\Sigma_z}{2}$ is the spin operator of the hidden-charm pentaquark states. At the hadron level, the axial charges read
		\begin{eqnarray}\left \langle
			P_{\psi}^{\Delta++}; \pi_{0}
			|
			\frac{d_{1}}{f_\pi}\bar{P}_{\psi}^{\Delta++}\frac{\Sigma_z}{2}\partial_{z} \pi_{0}P_{\psi}^{\Delta++}
			|
			P_{\psi}^{\Delta++}
			\right\rangle	=	\frac{d_{1}}{2}\frac{q_z}{f_\pi}.\label{equ:23}	
		\end{eqnarray}
		At the quark level, the axial charge of $P_{\psi}^{\Delta++}$ with the spin configuration $J^{P}=\frac{1}{2}^{-}(\frac{1}{2}^{+}\otimes0^{-})$ read
		\begin{eqnarray}
			\langle P_{\psi}^{\Delta++} , +\frac{1}{2} ;~\pi_0| \mathscr{L}_{quark}|P_{\psi}^{\Delta++},
			+\frac{1}{2}\rangle
			=\frac{2}{3}\frac{q_z}{f_\pi}g_q.\label{equ:24}
		\end{eqnarray}
By comparing the hadron-level and quark-level matrix elements for the nucleon and the pentaquark, we establish the relation:
		\begin{eqnarray}
			\frac{\frac{1}{2}g_{A}}{\frac{5}{6}g_{q}}	
			=
			\frac{\frac{d_{1}}{2}}{\frac{2}{3}g_{q}}.\label{equ:28}
		\end{eqnarray}
Thus, we obtain $d_{1}= \frac{4}{5}g_{A}$. Similarly, we can obtain the axial charge of the pentaquark states with other flavor-spin configurations. The Lagrangian of pentaquark state for $J^{P}=\frac{1}{2}^{-}(\frac{1}{2}^{+}\otimes1^{-})$ in $\mathbf{10}_f$ flavor representation reads
		\begin{eqnarray}
		\mathscr{L}^{\frac{1}{2}}_{2}
		&=\rm Tr\Big(d_{2}\bar{\mathcal{P}} \gamma_{\mu}\gamma^{5}\partial^{\mu}\Phi \mathcal{P}\Big),
	\end{eqnarray}	
	The Lagrangian of pentaquark state for $J^{P}=\frac{3}{2}^{-}(\frac{1}{2}^{+}\otimes1^{-})$ in $\mathbf{10}_f$ flavor representation reads
		\begin{eqnarray}
		\mathscr{L}^{\frac{3}{2}}_{3}
		&=\rm Tr\Big(d_{3}\bar{\mathcal{P}}^{\nu} \gamma_{\mu}\gamma^{5}\partial^{\mu}\Phi \mathcal{P}_{\nu}\Big).
	\end{eqnarray}	

		\renewcommand{\arraystretch}{1.5}
		\begin{table*}[htbp]
			\centering
			\caption{The axial charges of the decuplet hidden-charm pentaquark family in $\mathbf{10}_f$ flavor representation. The  $J_{b} ^{P_{b}}\otimes J_{m} ^{P_{m}} $  correspond to the angular momentum and parity of baryon and meson, respectively.}
			\label{tab_23}
			\vspace{0.5em}
			\resizebox{180mm}{!}{	
				\begin{tabular}{ c|c|c|c|c|c}
				\toprule[1.0pt]
				\toprule[1.0pt]
					Couplings & Coefficients &Wave functions & $J_{b} ^{P_{b}}\otimes J_{m} ^{P_{m}} $ & $I(J^{P})$ & Results
					\\
					\hline
					\multirow{3}{*}{$P_{\psi}^{\Delta++}{P_{\psi}^{\Delta++}}{\pi_{0}}$}
					&	$d_{1}$ 
					& \multirow{3}{*}{${P_{\psi}^{\Delta++}}:{\bar D^{(*)0}}\Sigma_c^{++}$}
					&	$\frac{1}{2}^{+}\times0^{-}$
					&	$\frac{1}{2}(\frac{1}{2})^{-}$
					&	$ \frac{4}{5}g_{A}$
					\\
					\cline{2-2}
					\cline{4-6}

					&	$d_{2}$
					&		
					&	\multirow{2}{*}{$\frac{1}{2}^{+}\times1^{-}$}
					&	$\frac{1}{2}(\frac{1}{2})^{-}$
					&	$\frac{7}{15}g_{A}$
					\\
					\cline{2-2}
					\cline{5-6}

					&	$d_{3}$	
					&
					&	
					&	$\frac{1}{2}(\frac{3}{2})^{-}$		
					&	$\frac{7}{5}g_{A}$
					\\
					\hline					
					
					\multirow{3}{*}{${P_{\psi}^{\Delta+}}{P_{\psi}^{\Delta+}}\pi_{0}$}
					&	$\frac{1}{3}d_{1}$
					&				\multirow{3}{*}{${P_{\psi}^{\Delta+}}: \sqrt{\frac{2}{3}}{\bar D^{(*)0}}\Sigma_c^{+}+\sqrt{\frac{1}{3}}{D^{(*)-}}\Sigma_c^{++}$}
					&	$\frac{1}{2}^{+}\times0^{-}$
					&	$\frac{1}{2}(\frac{1}{2})^{-}$
					&	$ \frac{4}{15}g_{A}$
					\\
					\cline{2-2}
					\cline{4-6}

					&	$\frac{1}{3}d_{2}$
					&		
					&	\multirow{2}{*}{$\frac{1}{2}^{+}\times1^{-}$}
					&	$\frac{1}{2}(\frac{1}{2})^{-}$
					&	$\frac{7}{45}g_{A}$
					\\
					\cline{2-2}
					\cline{5-6}

					&	$\frac{1}{3}d_{3}$
					&
					&	
					&	$\frac{1}{2}(\frac{3}{2})^{-}$
					&	$\frac{7}{15}g_{A}$
					\\

						\hline
					\multirow{3}{*}{$P_{\psi}^{\Delta0}{P_{\psi}^{\Delta0}}{\pi_{0}}$}
					&	$-\frac{1}{3}d_{1}$ 
					
					& \multirow{3}{*}{${P_{\psi}^{\Delta0}}:\sqrt{\frac{2}{3}}{ D^{(*)-}}\Sigma_c^{+}+\sqrt{\frac{1}{3}}{\bar D^{(*)0}}\Sigma_c^{0}$}
					&	$\frac{1}{2}^{+}\times0^{-}$
					&	$\frac{1}{2}(\frac{1}{2})^{-}$
					&	$- \frac{4}{15}g_{A}$
					\\
					\cline{2-2}
					\cline{4-6}

					&	$-\frac{1}{3}d_{2}$
					&		
					&	\multirow{2}{*}{$\frac{1}{2}^{+}\times1^{-}$}
					&	$\frac{1}{2}(\frac{1}{2})^{-}$
					&	$-\frac{7}{45}g_{A}$
					\\
					\cline{2-2}
					\cline{5-6}

					&	$-\frac{1}{3}d_{3}$	
					&
					&	
					&	$\frac{1}{2}(\frac{3}{2})^{-}$		
					&	$-\frac{7}{15}g_{A}$
					\\
					\hline					
					
					\multirow{3}{*}{${P_{\psi}^{\Delta-}}{P_{\psi}^{\Delta-}}\pi_{0}$}
					&	$-d_{1}$
					&				\multirow{3}{*}{${P_{\psi}^{\Delta-}}: { D^{(*)-}}\Sigma_c^{0}$}
					&	$\frac{1}{2}^{+}\times0^{-}$
					&	$\frac{1}{2}(\frac{1}{2})^{-}$
					&	$- \frac{4}{5}g_{A}$
					\\
					\cline{2-2}
					\cline{4-6}

					&	$-d_{2}$
					&		
					&	\multirow{2}{*}{$\frac{1}{2}^{+}\times1^{-}$}
					&	$\frac{1}{2}(\frac{1}{2})^{-}$
					&	$-\frac{7}{15}g_{A}$
					\\
					\cline{2-2}
					\cline{5-6}

					&	$-d_{3}$
					&
					&	
					&	$\frac{1}{2}(\frac{3}{2})^{-}$
					&	$-\frac{7}{5}g_{A}$
					\\
					\hline
					\multirow{3}{*}{${P_{\psi s}^{\Sigma+}}{P_{\psi s}^{\Sigma+}}\pi_{0}$}
					&	$\frac{2}{3}d_{1}$
					&	\multirow{3}{*}{${P_{\psi s}^{\Sigma+}}:\sqrt{\frac{2}{3}}{\bar D^{(*)0}}\Xi_c^{\prime +}+\sqrt{\frac{1}{3}}{ D_s^{(*)-}}\Sigma_c^{++}$}
					&	$\frac{1}{2}^{+}\times0^{-}$
					&	$1(\frac{1}{2})^{-}$
					&	$ \frac{8}{15}g_{A}$
					\\
					\cline{2-2}
					\cline{4-6}

					&	$\frac{2}{3}d_{2}$
					&		
					&	\multirow{2}{*}{$\frac{1}{2}^{+}\times1^{-}$}
					&	$1(\frac{1}{2})^{-}$
					&	$\frac{14}{15}g_{A}$
					\\
					\cline{2-2}
					\cline{5-6}

					&	$\frac{2}{3}d_{3}$
					&	
					&		
					&	$1(\frac{3}{2})^{-}$
					&	$\frac{14}{15}g_{A}$
					\\
					\hline
				
				\multirow{3}{*}{${P_{\psi s}^{\Sigma0}}{P_{\psi s}^{\Sigma0}}\pi_{0}$}
				&	$0$
				&	\multirow{3}{*}{${P_{\psi s}^{\Sigma0}}:\sqrt{\frac{1}{3}}({ D^{(*)-}}\Xi_c^{\prime +}+
					{\bar D^{(*)0}}\Xi_c^{\prime 0})+\sqrt{\frac{1}{3}}{D_s^{(*)-}}\Sigma_c^{+}$}
				&	$\frac{1}{2}^{+}\times0^{-}$
				&	$1(\frac{1}{2})^{-}$
				&	$0$
				\\
				\cline{2-2}
				\cline{4-6}

				&	$0$
				&		
				&	\multirow{2}{*}{$\frac{1}{2}^{+}\times1^{-}$}
				&	$1(\frac{1}{2})^{-}$
				&	$0$
				\\
				\cline{2-2}
				\cline{5-6}

				&	$0$
				&	
				&		
				&	$1(\frac{3}{2})^{-}$
				&	$0$
				\\
				\hline
					\multirow{3}{*}{${P_{\psi s}^{\Sigma^{-}}}{P_{\psi s}^{\Sigma^{-}}}\pi_{0}$}
					&	$-\frac{2}{3}d_{1}$
					&	\multirow{3}{*}{${P_{\psi s}^{\Sigma^{-}}}: \sqrt{\frac{2}{3}}{ D^{(*)-}}\Xi_c^{\prime 0}+\sqrt{\frac{1}{3}}{ D_s^{(*)-}}\Sigma_c^{0}$}
					&	$\frac{1}{2}^{+}\times0^{-}$
					&	$1(\frac{1}{2})^{-}$
					&	$- \frac{8}{15}g_{A}$
					\\
					\cline{2-2}
					\cline{4-6}

					&	$-\frac{2}{3}d_{2}$
					&	
					&	\multirow{2}{*}{$\frac{1}{2}^{+}\times1^{-}$}
					&	$1(\frac{1}{2})^{-}$
					&	$-\frac{14}{45}g_{A}$
					\\
					\cline{2-2}
					\cline{5-6}

					&	$-\frac{2}{3}d_{3}$
					&	
					&	
					&	$1(\frac{3}{2})^{-}$
					&	$-\frac{14}{15}g_{A}$
					\\
					\hline
					
					\hline
				\multirow{3}{*}{${P_{\psi ss}^{N0}}{P_{\psi ss}^{N0}}\pi_{0}$}
				&	$\frac{1}{3}d_{1}$
				&	\multirow{3}{*}{${P_{\psi ss}^{N0}}:\sqrt{\frac{1}{3}}{\bar D^{(*)0}}\Omega_c^{0}+\sqrt{\frac{2}{3}}{D_s^{(*)-}}\Xi_c^{\prime +}$}
				&	$\frac{1}{2}^{+}\times0^{-}$
				&	$1(\frac{1}{2})^{-}$
				&	$ \frac{4}{15}g_{A}$
				\\
				\cline{2-2}
				\cline{4-6}

				&	$\frac{1}{3}d_{2}$
				&		
				&	\multirow{2}{*}{$\frac{1}{2}^{+}\times1^{-}$}
				&	$1(\frac{1}{2})^{-}$
				&	$\frac{7}{45}g_{A}$
				\\
				\cline{2-2}
				\cline{5-6}

				&	$\frac{1}{3}d_{3}$
				&	
				&		
				&	$1(\frac{3}{2})^{-}$
				&	$\frac{7}{15}g_{A}$
				\\
				\hline
					\multirow{3}{*}{${P_{\psi ss}^{N^{-}}}{P_{\psi ss}^{N^{-}}}\pi_{0}$}
					&	$-\frac{1}{3}d_{1}$
					&	\multirow{3}{*}{${P_{\psi ss}^{N^{-}}}: \sqrt{\frac{1}{3}}{D^{(*)-}}\Omega_c^{0}+\sqrt{\frac{2}{3}}{ D_s^{(*)-}}\Xi_c^{\prime 0}$}
					&	$\frac{1}{2}^{+}\times0^{-}$
					&	$\frac{1}{2}(\frac{1}{2})^{-}$
					&	$- \frac{4}{15}g_{A}$
					\\
					\cline{2-2}
					\cline{4-6}

					&	$-\frac{1}{3}d_{2}$
					&	
					&	\multirow{2}{*}{$\frac{1}{2}^{+}\times1^{-}$}
					&	$\frac{1}{2}(\frac{1}{2})^{-}$
					&	$-\frac{7}{45}g_{A}$
					\\
					\cline{2-2}
					\cline{5-6}

					&	$-\frac{1}{3}d_{3}$
					&	
					&	
					&	$\frac{1}{2}(\frac{3}{2})^{-}$
					&	$-\frac{7}{15}g_{A}$\\
					\hline
					\multirow{3}{*}{${P_{\psi sss}^{\Lambda -}}{P_{\psi sss}^{\Lambda -}}\pi_{0}$}
					&	$0$
					&	\multirow{3}{*}{${P_{\psi sss}^{\Lambda -}}: {D_s^{(*)-}}\Omega_c^{0}$}
					&	$\frac{1}{2}^{+}\times0^{-}$
					&	$\frac{1}{2}(\frac{1}{2})^{-}$
					&	$0$
					\\
					\cline{2-2}
					\cline{4-6}

					&	$0$
					&	
					&	\multirow{2}{*}{$\frac{1}{2}^{+}\times1^{-}$}
					&	$\frac{1}{2}(\frac{1}{2})^{-}$
					&	$0$
					\\
					\cline{2-2}
					\cline{5-6}

					&	$0$
					&	
					&	
					&	$\frac{1}{2}(\frac{3}{2})^{-}$
					&	$0$
					\\
				\toprule[1.0pt]
				\toprule[1.0pt]				
			\end{tabular}}	
		\end{table*}	
	
		Similarly, we can also obtain the coupling constants $ d_{2}= \frac{7}{15}g_{A}$ and $ d_{3}= \frac{7}{5}g_{A}$. The axial charges of the decuplet hidden-charm molecular family in $\mathbf{10}_f$ flavor representation are listed in Table \ref{tab_23}. As presented in Table \ref{tab_23}, the axial charges of the decuplet hidden-charm pentaquark family are comparable to that of the nucleon, and generally larger compared to that of the octet hidden-charm pentaquark family as shown in our previous work in Ref.~\cite{Li:2024wxr}. We also notice that the axial charges of $P_{\psi s}^{\Sigma0}$ and ${P_{\psi sss}^{\Lambda -}}$ in $\mathbf{10}_f$ flavor representation are zero. 
	
	At the hadronic level, the axial charge serves as a discriminating marker for the flavor-spin structure of pentaquark states. The resulting variation in its calculated value, within the quark-hadron dual picture, provides a robust basis for elucidating the strong decay patterns of structurally distinct pentaquarks in future research. The tabulated axial charges in Table \ref{tab_23} are thus poised to play a crucial role in guiding subsequent calculations within the framework of chiral perturbation theory (details in our previous work in Ref.~\cite{Li:2024jlq}).

	\section{Magnetic Moments of The decuplet Hidden-Charm Molecular Pentaquark Family }
\label{sec4}
Within the framework of the constituent quark model, we calculate the magnetic moments of hadrons. The quark model has been extensively applied over the past decades to investigate a wide range of hadronic properties, with the magnetic moments of hadronic molecular states being a recent focus of interest~\cite{Karliner:2015ina,Nakamura:2021dix,Ling:2021lmq}. The total magnetic moment of a hadron can be decomposed into two components: the spin magnetic moment ($\mu_{spin}$) and the orbital magnetic moment ($\mu_{orbital}$)
\begin{eqnarray}
	\mu = \mu_{spin}+ \mu_{orbital}.
\end{eqnarray}
In this work, we investigate the magnetic moments of $ S $-wave pentaquark states within the molecular picture, where they are considered as bound states of a singly charmed baryon and an anti-charmed meson. Accordingly, the orbital magnetic moment $\mu_{orbital}$ arising from the relative motion between the hadronic constituents is omitted from our subsequent calculations. For the spin magnetic moment ($\mu_{spin}$), the operators of the magnetic moments at the quark level are
\begin{eqnarray}
	\hat{\mu}_{spin} = {\sum_{i}}{\frac{Q_{i}}{2M_{i}}}\hat{\sigma}_{i},
\end{eqnarray}
where $Q_i$, $M_i $, and $\hat{\sigma}_{i}$ represent charge, mass and Pauli spin matrix of the $i$th quark, respectively. For hidden-charm molecular pentaquark states, the magnetic moment of an S-wave molecular pentaquark state is composed of the baryon spin magnetic moment and the meson spin magnetic moment
\begin{eqnarray}
	\hat{\mu} = \hat{\mu}_{B} + \hat{\mu}_{M},
\end{eqnarray}
Here, the subscripts $B$ and $M$ denote the baryon and meson components, respectively. Table \ref{tab_5} summarizes the expressions for the magnetic moments of the singly charmed baryons and anti-charmed mesons, where each is given as a sum over the magnetic moments of their constituent quarks. For the numerical analysis, the constituent quark masses are used as the primary input parameters \cite{Wang:2018gpl}:
\begin{align}
	m_u \ &=\ m_d \ =\  0.336\ \mbox{GeV}, \nonumber
	\\ 	
	m_s \ &=\ 0.540\ \mbox{GeV},\nonumber
	\\
	m_c \ &=\ 1.660\ \mbox{GeV}. \nonumber	
\end{align}

\renewcommand\tabcolsep{0.35cm}
\renewcommand{\arraystretch}{1.50}
\begin{table}[t]
	\centering
	\vspace{0.5em}
	\caption{The magnetic moments of the singly charmed baryons and anti-charmed mesons, in unit of the nuclear magnetic moment $\mu_{N}$.}
	\label{tab_5}
	\begin{tabular}{c|c|c|c}
		\toprule[1.0pt]
		\toprule[1.0pt]
		Flavor & Hadrons & Expressions &  Results
		\\
		\hline
		
		\multirow{6}{*}{$6_f$}	
		&	$\Sigma_{c}^{++}$
		&	${\frac{4}{3}}\mu_{u} - {\frac{1}{3}}\mu_{c}$
		&   $2.36$
		\\

		&	$\Sigma_{c}^{+}$
		&	${\frac{2}{3}}\mu_{u} + {\frac{2}{3}}\mu_{d} - {\frac{1}{3}}\mu_{c}$
		&   $0.49$
		\\

		&	$\Sigma_{c}^{0}$
		&	${\frac{4}{3}}\mu_{d} - {\frac{1}{3}}\mu_{c}$
		&   $-1.37$
		\\

		&	$\Xi_{c}^{\prime +}$
		&	${\frac{2}{3}}\mu_{u} + {\frac{2}{3}}\mu_{s} - {\frac{1}{3}}\mu_{c}$
		&   $0.73$
		\\

		&	$\Xi_{c}^{\prime 0}$
		&	${\frac{2}{3}}\mu_{d} + {\frac{2}{3}}\mu_{s} - {\frac{1}{3}}\mu_{c}$
		&   $-1.13$
		\\

		&	$\Omega_{c}^{0}$
		&	${\frac{4}{3}}\mu_{s}-{\frac{1}{3}}\mu_{c}$
		&   $-0.90$
		\\
		\hline
		
		\multirow{3}{*}{$\bar{3}_f$}	
		&	$\Xi_{c}^{+}$
		&	$\mu_{c}$
		&   $0.38$
		\\
		
		&	$\Xi_{c}^{0}$
		&	$\mu_{c}$
		&   $0.38$
		\\

		&	$\Lambda_{c}^{+}$
		&	$\mu_{c}$
		&   $0.38$
		\\
		\hline
		
		\multirow{3}{*}{$3_f$}	
		&	$\bar{D}^{*0}$
		&	$\mu_{u} + \mu_{\bar{c}}$
		&   $1.48$
		\\

		&	$D^{*-}$
		&	$\mu_{d} + \mu_{\bar{c}}$
		&   $-1.31$
		\\

		&	$D_{s}^{*-}$
		&	$\mu_{s} + \mu_{\bar{c}}$
		&   $-0.96$
		\\
		\toprule[1.0pt]
		\toprule[1.0pt]
	\end{tabular}
\end{table}

\begin{table*}[htbp]
	\centering
	\caption{The magnetic moments of the $ S $-wave $\frac{1}{2}^{+}\otimes0^{-}=\frac{1}{2}^{-}$ decuplet hidden-charm pentaquark states  (in unit of $\mu_{N}$).}
	\label{tab_6}
	\vspace{0.5em}
	\begin{tabular}{c|c|c}
		\toprule[1.0pt]
	\toprule[1.0pt]
		$\frac{1}{2}^{+}\otimes0^{-}=\frac{1}{2}^{-}$ States &Expressions & Results\tabularnewline
	
		\hline
	$P_{\psi}^{\Delta++}$ & $\mu_{\Sigma_{c}^{++}}$ & 2.36\tabularnewline
\hline
$P_{\psi}^{\Delta+}$ & $\frac{2}{3}\mu_{\Sigma_{c}^{+}}+\frac{1}{3}\mu_{\Sigma_{c}^{++}}$ & 1.12\tabularnewline
\hline
$P_{\psi}^{\Delta0}$ & $\frac{2}{3}\mu_{\Sigma_{c}^{+}}+\frac{1}{3}\mu_{\Sigma_{c}^{0}}$ & -0.13\tabularnewline
\hline
$P_{\psi}^{\Delta-}$ & $\mu_{\Sigma_{c}^{0}}$ & -1.37\tabularnewline
\hline
$P_{\psi s}^{\Sigma+}$ & $\frac{2}{3}\mu_{\Xi_{c}^{\prime+}}+\frac{1}{3}\mu_{\Sigma_{c}^{++}}$ & 1.27\tabularnewline
\hline
$P_{\psi s}^{\Sigma0}$ & $\frac{1}{3}\mu_{\Xi_{c}^{\prime+}}+\frac{1}{3}\mu_{\Xi_{c}^{\prime0}}+\frac{1}{3}\mu_{\Sigma_{c}^{+}}$ & 0.03\tabularnewline
\hline
$P_{\psi s}^{\Sigma-}$ & $\frac{2}{3}\mu_{\Xi_{c}^{\prime0}}+\frac{1}{3}\mu_{\Sigma_{c}^{0}}$ & -1.21\tabularnewline
\hline
$P_{\psi ss}^{N0}$ & $\frac{2}{3}\mu_{\Xi_{c}^{\prime+}}+\frac{1}{3}\mu_{\Omega_{c}^{0}}$ & 0.19\tabularnewline
\hline
$P_{\psi ss}^{N-}$ & $\frac{2}{3}\mu_{\Xi_{c}^{\prime0}}+\frac{1}{3}\mu_{\Omega_{c}^{0}}$ & -1.05\tabularnewline
\hline
$P_{\psi sss}^{\Lambda-}$ & $\mu_{\Omega_{c}^{0}}$ & -0.90\tabularnewline	
		\toprule[1.0pt]
	\toprule[1.0pt]
	\end{tabular}
\end{table*}

\begin{table*}[htbp]
	\centering
	\caption{The magnetic moments of the $ S $-wave $\frac{1}{2}^{+}\otimes1^{-}=\frac{1}{2}^{-}$ decuplet hidden-charm pentaquark states  (in unit of $\mu_{N}$).}
	\label{tab_7}
	\vspace{0.5em}
	\begin{tabular}{c|c|c}
		\toprule[1.0pt]
		\toprule[1.0pt]
		$\frac{1}{2}^{+}\otimes1^{-}=\frac{1}{2}^{-}$ States &Expressions & Results\tabularnewline
		
		\hline
	$P_{\psi}^{\Delta++}$ & $\frac{2}{3}\mu_{\bar{D}^{*0}}-\frac{1}{3}\mu_{\Sigma_{c}^{++}}$ & 0.20\tabularnewline
	\hline
	$P_{\psi}^{\Delta+}$ & $\frac{4}{9}\mu_{\bar{D}^{*0}}+\frac{2}{9}\mu_{D^{*-}}-\frac{2}{9}\mu_{\Sigma_{c}^{+}}-\frac{1}{9}\mu_{\Sigma_{c}^{++}}$ & -0.002\tabularnewline
	\hline
	$P_{\psi}^{\Delta0}$ & $\frac{4}{9}\mu_{D^{*-}}+\frac{2}{9}\mu_{\bar{D}^{*0}}-\frac{2}{9}\mu_{\Sigma_{c}^{+}}-\frac{1}{9}\mu_{\Sigma_{c}^{0}}$ & -0.21\tabularnewline
	\hline
	$P_{\psi}^{\Delta-}$ & $\frac{2}{3}\mu_{D^{*-}}-\frac{1}{3}\mu_{\Sigma_{c}^{0}}$ & -0.42\tabularnewline
	\hline
	$P_{\psi s}^{\Sigma+}$ & $\frac{4}{9}\mu_{\bar{D}^{*0}}+\frac{2}{9}\mu_{D_{s}^{*-}}-\frac{2}{9}\mu_{\Xi_{c}^{\prime+}}-\frac{1}{9}\mu_{\Sigma_{c}^{++}}$ & 0.02\tabularnewline
	\hline
	$P_{\psi s}^{\Sigma0}$ & $\frac{2}{9}\mu_{D^{*-}}-\frac{1}{9}\mu_{\Xi_{c}^{\prime+}}+\frac{2}{9}\mu_{\bar{D}^{*0}}-\frac{1}{9}\mu_{\Xi_{c}^{\prime0}}+\frac{2}{9}\mu_{D_{s}^{*-}}-\frac{1}{9}\mu_{\Sigma_{c}^{+}}$ & -0.18\tabularnewline
	\hline
	$P_{\psi s}^{\Sigma-}$ & $\frac{4}{9}\mu_{D^{*-}}+\frac{2}{9}\mu_{D_{s}^{*-}}-\frac{2}{9}\mu_{\Xi_{c}^{\prime0}}-\frac{1}{9}\mu_{\Sigma_{c}^{0}}$ & -0.39\tabularnewline
	\hline
	$P_{\psi ss}^{N0}$ & $\frac{4}{9}\mu_{D_{s}^{*-}}+\frac{2}{9}\mu_{\bar{D}^{*0}}-\frac{2}{9}\mu_{\Xi_{c}^{\prime+}}-\frac{1}{9}\mu_{\Omega_{c}^{0}}$ & -0.16\tabularnewline
	\hline
	$P_{\psi ss}^{N-}$ & $\frac{4}{9}\mu_{D_{s}^{*-}}+\frac{2}{9}\mu_{D^{*-}}-\frac{2}{9}\mu_{\Xi_{c}^{\prime0}}-\frac{1}{9}\mu_{\Omega_{c}^{0}}$ & -0.36\tabularnewline
	\hline
	$P_{\psi sss}^{\Lambda-}$ & $\frac{2}{3}\mu_{D_{s}^{*-}}-\frac{1}{3}\mu_{\Omega_{c}^{0}}$ & -0.34\tabularnewline
		\toprule[1.0pt]
		\toprule[1.0pt]
	\end{tabular}
\end{table*}

\begin{table*}[htbp]
	\centering
	\caption{The magnetic moments of the $ S $-wave 	$1^{-}\otimes\frac{1}{2}^{+}=\frac{3}{2}^{-}$ decuplet hidden-charm pentaquark states (in unit of $\mu_{N}$).}
	\label{tab_8}
	\vspace{0.5em}
	\begin{tabular}{c|c|c}
		\toprule[1.0pt]
		\toprule[1.0pt]
	$1^{-}\otimes\frac{1}{2}^{+}=\frac{3}{2}^{-}$ States &Expressions & Results\tabularnewline
		
		\hline
	$P_{\psi}^{\Delta++}$ & $\mu_{\bar{D}^{*0}}+\mu_{\Sigma_{c}^{++}}$ & 3.84\tabularnewline
	\hline
	$P_{\psi}^{\Delta+}$ & $\frac{2}{3}\mu_{\bar{D}^{*0}}+\frac{2}{3}\mu_{\Sigma_{c}^{+}}+\frac{1}{3}\mu_{D^{*-}}+\frac{1}{3}\mu_{\Sigma_{c}^{++}}$ & 1.67\tabularnewline
	\hline
	$P_{\psi}^{\Delta0}$ & $\frac{2}{3}\mu_{D^{*-}}+\frac{2}{3}\mu_{\Sigma_{c}^{+}}+\frac{1}{3}\mu_{\bar{D}^{*0}}+\frac{1}{3}\mu_{\Sigma_{c}^{0}}$ & -0.50\tabularnewline
	\hline
	$P_{\psi}^{\Delta-}$ & $\mu_{D^{*-}}+\mu_{\Sigma_{c}^{0}}$ & -2.67\tabularnewline
	\hline
	$P_{\psi s}^{\Sigma+}$ & $\frac{2}{3}\mu_{\bar{D}^{*0}}+\frac{2}{3}\mu_{\Xi_{c}^{\prime+}}+\frac{1}{3}\mu_{D_{s}^{*-}}+\frac{1}{3}\mu_{\Sigma_{c}^{++}}$ & 1.94\tabularnewline
	\hline
	$P_{\psi s}^{\Sigma0}$ & $\frac{1}{3}\mu_{D^{*-}}+\frac{1}{3}\mu_{\bar{D}^{*0}}+\frac{1}{3}\mu_{D_{s}^{*-}}+\frac{1}{3}\mu_{\Xi_{c}^{\prime+}}+\frac{1}{3}\mu_{\Xi_{c}^{\prime0}}+\frac{1}{3}\mu_{\Sigma_{c}^{+}}$ & -0.23\tabularnewline
	\hline
	$P_{\psi s}^{\Sigma-}$ & $\frac{2}{3}\mu_{D^{*-}}+\frac{2}{3}\mu_{\Xi_{c}^{\prime0}}+\frac{1}{3}\mu_{D_{s}^{*-}}+\frac{1}{3}\mu_{\Sigma_{c}^{0}}$ & -2.40\tabularnewline
	\hline
	$P_{\psi ss}^{N0}$ & $\frac{2}{3}\mu_{D_{s}^{*-}}+\frac{2}{3}\mu_{\Xi_{c}^{\prime+}}+\frac{1}{3}\mu_{\bar{D}^{*0}}+\frac{1}{3}\mu_{\Omega_{c}^{0}}$ & 0.04\tabularnewline
	\hline
	$P_{\psi ss}^{N-}$ & $\frac{2}{3}\mu_{D_{s}^{*-}}+\frac{2}{3}\mu_{\Xi_{c}^{\prime0}}+\frac{1}{3}\mu_{D^{*-}}+\frac{1}{3}\mu_{\Omega_{c}^{0}}$ & -2.13\tabularnewline
	\hline
	$P_{\psi sss}^{\Lambda-}$ & $\mu_{D_{s}^{*-}}+\mu_{\Omega_{c}^{0}}$ & -1.85\tabularnewline
		\toprule[1.0pt]
		\toprule[1.0pt]
	\end{tabular}
\end{table*}		

Building on the magnetic moments of the singly charmed baryons and anti-charmed mesons, we derive the corresponding moments for the decuplet hidden-charm molecular pentaquark states. The resulting analytical expressions and their corresponding numerical values for the S-wave $\frac{1}{2}^{+}\otimes0^{-}=\frac{1}{2}^{-}$, $\frac{1}{2}^{+}\otimes1^{-}=\frac{1}{2}^{-}$ and $\frac{1}{2}^{+}\otimes1^{-}=\frac{3}{2}^{-}$ pentaquark family in $\mathbf{10}_{f}$ representation are summarized in Tables \ref{tab_6}, \ref{tab_7} and \ref{tab_8}. Thus, we can summarize several points:

\begin{itemize}
	\item The magnetic moments of the S-wave decuplet hidden-charm pentaquark states are strongly dependent on their specific flavor and spin configurations. Notably, the magnetic moments of the $\frac{1}{2}^{+} \otimes 1^{-} = \frac{1}{2}^{-}$ states are significantly smaller than those of the $\frac{1}{2}^{+} \otimes 0^{-} = \frac{1}{2}^{-}$ and $\frac{1}{2}^{+} \otimes 1^{-} = \frac{3}{2}^{-}$ configurations. Furthermore, for the S-wave $\frac{1}{2}^{+} \otimes 0^{-} = \frac{1}{2}^{-}$ and $\frac{1}{2}^{+} \otimes 1^{-} = \frac{3}{2}^{-}$ decuplet hidden-charm pentaquark states, the magnetic moments are approximately proportional to their electric charges.
	\item 	 The magnetic moments of decuplet pentaquark states sharing the same spin-flavor configuration satisfy the following Hao-Song sum rules
	\begin{eqnarray}
		&&	\mu_{P_{\psi}^{\Delta+}}-
		\mu_{P_{\psi}^{\Delta0}}+\mu_{P_{\psi s}^{\Sigma-}}-
		\mu_{P_{\psi s}^{\Sigma+}}
		+\mu_{P_{\psi ss}^{N0}}-
		\mu_{P_{\psi ss}^{N-}}=0, \nonumber\\
		&&	\mu_{P_{\psi}^{\Delta++}}+\mu_{P_{\psi}^{\Delta-}}
		+\mu_{P_{\psi sss}^{\Lambda-}}=
		3\mu_{P_{\psi s}^{\Sigma0}}. 
	\end{eqnarray}
	Formulated similarly to the Coleman-Glashow sum rule for octet baryon magnetic moments in the SU(3) flavor symmetry limit \cite{Coleman:1961jn,Cheng:1997kr,Dahiya:2002fp}, the first Hao-Song sum rule is shown to be valid for decuplet pentaquark states of corresponding spin-flavor configurations. The second sum rule further delineates the relationships governing the remaining decuplet pentaquark magnetic moments. These empirical rules furnish novel perspectives for exploring the substructure of hidden-charm pentaquarks. Our predictions, established in the absence of coupled-channel effects, will therefore be instrumental in guiding subsequent research into their internal composition.

\end{itemize}

	\section{SUMMARY}
	\label{sec5}
The systematic study of exotic hadrons continues to provide critical tests for our understanding of non-perturbative QCD. In this context, the axial charge and the magnetic moment stand out as fundamental observables that probe distinct aspects of hadronic structure: the former is intimately connected to chiral symmetry and spin content, while the latter reveals the spatial distribution of internal currents and charges.

In this work, we have performed a comprehensive analysis of the decuplet hidden-charm molecular pentaquarks within the quark model. Our calculation of the axial charges demonstrates that their magnitudes are comparable to that of the nucleon, suggesting significant quark spin contributions. These results provide essential input for constructing effective theories, such as chiral perturbation theory, for these exotic states.

Regarding the magnetic moments, our findings reveal a rich and structured pattern. These magnetic moments are highly sensitive to the specific flavor and spin configuration of the pentaquark. Notably, the magnetic moments for the $\frac{1}{2}^{+} \otimes 0^{-} = \frac{1}{2}^{-}$ and $\frac{1}{2}^{+} \otimes 1^{-} = \frac{3}{2}^{-}$ states are found to be approximately proportional to their electric charges. A key outcome of this work is the derivation of the Hao-Song sum rules, a set of relations obeyed by the magnetic moments of decuplet pentaquarks sharing the same spin-flavor structure, which persist even under SU(3) symmetry breaking.

As fundamental physical observables, axial charges and magnetic moments offer valuable pathways toward experimental characterization. Although pentaquark states are short-lived, their magnetic moments could, in principle, be accessed via Thomas precession in a uniform magnetic field. Axial couplings, on the other hand, can be more directly extracted from the decay widths of strong decays. We anticipate that the predictions presented here-for both axial charges and magnetic moments-will serve as a valuable benchmark for future experimental analyses and theoretical studies, thereby advancing our understanding of the molecular nature and internal dynamics of multiquark systems.

		\vspace{1em}
	\section*{Acknowledgments}
	This project is supported by the National Natural Science Foundation of China under
	Grants No. 11905171. This work is also supported by Natural Science Basic Research
	Program of Shaanxi (Program No. 2025JC-YBMS-039) and Young Talent
	Fund of Xi'an Association for Science and Technology (Grant No. 959202413087).


\begin{thebibliography}{00}
		
		\bibitem{Gell-Mann:1964ewy}
		M.~Gell-Mann,
		``A Schematic Model of Baryons and Mesons,''
		Phys. Lett. \textbf{8} (1964) 214-215.
		
		
		
		\bibitem{LHCb:2016axx}
		R.~Aaij \textit{et al.} [LHCb],
		``Observation of $J/\psi\phi$ structures consistent with exotic states from amplitude analysis of $B^+\to J/\psi \phi K^+$ decays,''
		Phys. Rev. Lett. \textbf{118} (2017) no.2, 022003.
		
		\bibitem{BESIII:2016adj}
		M.~Ablikim \textit{et al.} [BESIII],
		``Evidence of Two Resonant Structures in $e^+ e^- \to \pi^+ \pi^- h_c$,''
		Phys. Rev. Lett. \textbf{118} (2017) no.9, 092002.
		
		\bibitem{LHCb:2018oeg}
		R.~Aaij \textit{et al.} [LHCb],
		``Evidence for an $\eta _c(1S) \pi ^-$ resonance in $B^0 \rightarrow \eta _c(1S) K^+\pi ^-$ decays,''
		Eur. Phys. J. C \textbf{78} (2018) no.12, 1019.
		
		\bibitem{LHCb:2015yax}
		R.~Aaij \textit{et al.} [LHCb],
		``Observation of $J/\psi p$ Resonances Consistent with Pentaquark States in $\Lambda_b^0 \to J/\psi K^- p$ Decays,''
		Phys. Rev. Lett. \textbf{115} (2015), 072001.
		
		\bibitem{LHCb:2019kea}
		R.~Aaij \textit{et al.} [LHCb],
		``Observation of a narrow pentaquark state, $P_c(4312)^+$, and of two-peak structure of the $P_c(4450)^+$,''
		Phys. Rev. Lett. \textbf{122} (2019) no.22, 222001.
		
		\bibitem{LHCb:2020jpq}
		R.~Aaij \textit{et al.} [LHCb],
		``Evidence of a $J/\psi\Lambda$ structure and observation of excited $\Xi^-$ states in the $\Xi^-_b \to J/\psi\Lambda K^-$ decay,''
		Sci. Bull. \textbf{66} (2021), 1278-1287.
		
	\bibitem{LHCb:2021chn}
	R.~Aaij \textit{et al.} [LHCb],
	``Evidence for a new structure in the $J/\psi p$ and $J/\psi \bar{p}$ systems in $B_s^0 \to J/\psi p \bar{p}$ decays,''
	Phys. Rev. Lett. \textbf{128} (2022) no.6, 062001.
	
	
	
	
	\bibitem{LHCb:2022ogu}
	R.~Aaij \textit{et al.} [LHCb],
	``Observation of a J/\ensuremath{\psi}\ensuremath{\Lambda} Resonance Consistent with a Strange Pentaquark Candidate in B-\textrightarrow{}J/\ensuremath{\psi}\ensuremath{\Lambda}p\textasciimacron{} Decays,''
	Phys. Rev. Lett. \textbf{131} (2023) no.3, 031901.
		
		\bibitem{Belle:2025rso}
		I.~Adachi \textit{et al.} [Belle and Belle-II],
		Phys. Rev. Lett. \textbf{135} (2025) no.4, 041901.	
		
		\bibitem{Gershon:2022xnn}
		T.~Gershon [LHCb],
		``Exotic hadron naming convention,''
		[arXiv:2206.15233 [hep-ex]].
		
		
		\bibitem{Deng:2022vkv}
		C.~R.~Deng,
		``Compact hidden charm pentaquark states and QCD isomers,''
		Phys. Rev. D \textbf{105} (2022) no.11, 116021.
		
		\bibitem{Wang:2019nvm}
		B.~Wang, L.~Meng and S.~L.~Zhu,
		``Spectrum of the strange hidden charm molecular pentaquarks in chiral effective field theory,''
		Phys. Rev. D \textbf{101} (2020) no.3, 034018.
		
		\bibitem{Xiao:2019gjd}
		C.~W.~Xiao, J.~Nieves and E.~Oset,
		``Prediction of hidden charm strange molecular baryon states with heavy quark spin symmetry,''
		Phys. Lett. B \textbf{799} (2019), 135051.
		
		\bibitem{Liu:2020hcv}
		M.~Z.~Liu, Y.~W.~Pan and L.~S.~Geng,
		``Can discovery of hidden charm strange pentaquark states help determine the spins of $P_c(4440)$ and $P_c(4457)$ ?,''
		Phys. Rev. D \textbf{103} (2021) no.3, 034003.	
		
		\bibitem{Peng:2020hql}
		F.~Z.~Peng, M.~J.~Yan, M.~S\'anchez S\'anchez and M.~P.~Valderrama,
		``The $P_{cs}(4459)$ pentaquark from a combined effective field theory and phenomenological perspective,''
		Eur. Phys. J. C \textbf{81} (2021) no.7, 666.
		
		\bibitem{Zhu:2021lhd}
		J.~T.~Zhu, L.~Q.~Song and J.~He,
		``$P_{cs}(4459)$ and other possible molecular states from $\Xi_{c}^{(*)}\bar{D}^{(*)}$ and $\Xi'_c\bar{D}^{(*)}$ interactions,''
		Phys. Rev. D \textbf{103} (2021) no.7, 074007.
		
		\bibitem{Du:2021bgb}
		M.~L.~Du, Z.~H.~Guo and J.~A.~Oller,
		``Insights into the nature of the Pcs(4459),''
		Phys. Rev. D \textbf{104} (2021) no.11, 114034.
		
		\bibitem{Chen:2021tip}
		R.~Chen,
		``Strong decays of the newly $P_{cs}(4459)$ as a strange hidden-charm $\Xi _c{\bar{D}}^*$ molecule,''
		Eur. Phys. J. C \textbf{81} (2021) no.2, 122.
		
		\bibitem{Yang:2021pio}
		F.~Yang, Y.~Huang and H.~Q.~Zhu,
		``Strong decays of the $P_{cs}(4459)$ as a $\Xi_c\bar{D}^{*}$ molecule,''
		Sci. China Phys. Mech. Astron. \textbf{64} (2021) no.12, 121011.
		
	\bibitem{Guo:2017jvc}
	F.~K.~Guo, C.~Hanhart, U.~G.~Mei{\ss}ner, Q.~Wang, Q.~Zhao and B.~S.~Zou,
	``Hadronic molecules,''
	Rev. Mod. Phys. \textbf{90} (2018) no.1, 015004
	[erratum: Rev. Mod. Phys. \textbf{94} (2022) no.2, 029901].
	
\bibitem{Dong:2021juy}
X.~K.~Dong, F.~K.~Guo and B.~S.~Zou,
``A survey of heavy-antiheavy hadronic molecules,''
Progr. Phys. \textbf{41} (2021), 65-93.


\bibitem{Yang:2024okq}
Z.~Y.~Yang, Q.~Wang and W.~Chen,
``Mass spectra of strange double charm pentaquarks with strangeness S={\ensuremath{-}}1,''
Phys. Rev. D \textbf{110} (2024) no.5, 056022.

\bibitem{Roca:2025zyi}
L.~Roca, J.~Song and E.~Oset,
``Study of hidden-charm, doubly-strange pentaquarks in $\Lambda_{b}\to J/\Psi \Xi^- K^+$ and $\Xi_b\to J/\Psi \Xi^- \pi^+$,''
[arXiv:2509.19840 [hep-ph]].

\bibitem{Liu:2025slt}
X.~Liu, Y.~Tan, X.~Chen, D.~Chen, H.~Huang and J.~Ping,
``Study on the properties of hidden-charm pentaquarks with double strangeness,''
Phys. Rev. D \textbf{112} (2025) no.1, 014036.


\bibitem{Clymton:2025zer}
S.~Clymton, H.~C.~Kim and T.~Mart,
``Double-strangeness hidden-charm pentaquarks,''
Phys. Rev. D \textbf{112} (2025) no.3, 034015.



		\bibitem{UCNA:2012fhw}
		M.~P.~Mendenhall \textit{et al.} [UCNA],
		``Precision measurement of the neutron $\beta$-decay asymmetry,''
		Phys. Rev. C \textbf{87} (2013) no.3, 032501.
		
		\bibitem{Mund:2012fq}
		D.~Mund, B.~Maerkisch, M.~Deissenroth, J.~Krempel, M.~Schumann, H.~Abele, A.~Petoukhov 	and T.~Soldner,
		``Determination of the Weak Axial Vector Coupling from a Measurement of the Beta-Asymmetry Parameter A in Neutron Beta Decay,''
		Phys. Rev. Lett. \textbf{110} (2013), 172502.
		
		\bibitem{Goldberger:1958tr}
		M.~L.~Goldberger and S.~B.~Treiman,
		``Decay of the pi meson,''
		Phys. Rev. \textbf{110} (1958), 1178-1184.
		
			\bibitem{Bijnens:1984ec}
		J.~Bijnens, H.~Sonoda and M.~B.~Wise,
		``On the Validity of Chiral Perturbation Theory for K0 anti-K0 Mixing,''
		Phys. Rev. Lett. \textbf{53} (1984), 2367.
		
		
		
		
		\bibitem{Jenkins:1991es}
		E.~E.~Jenkins and A.~V.~Manohar,
		``Chiral corrections to the baryon axial currents,''
		Phys. Lett. B \textbf{259}  (1991), 353-358.
		

		\bibitem{Zhu:2002tn}
		S.~L.~Zhu, G.~Sacco and M.~J.~Ramsey-Musolf,
		``Recoil order chiral corrections to baryon octet axial currents and large N(c) QCD,''
		Phys. Rev. D \textbf{66} (2002), 034021.
		
		\bibitem{Choi:2010ty}
		K.~S.~Choi, W.~Plessas and R.~F.~Wagenbrunn,
		``Axial charges of octet and decuplet baryons,''
		Phys. Rev. D \textbf{82} (2010), 014007.
		
		
		
		
		\bibitem{Dahiya:2023izc}
		H.~Dahiya, S.~Dutt, A.~Kumar and M.~Randhawa,
		``Axial-vector charges of the spin $\frac{1}{2}^+$ and spin $\frac{3}{2}^+$ light and charmed baryons in the SU(4) chiral quark constituent model,''
		Eur. Phys. J. Plus \textbf{138}, no.5, 441 (2023).
		\bibitem{Gao:2021hmv}
		F.~Gao and H.~S.~Li,
		``Magnetic moments of hidden-charm strange pentaquark states*,''
		Chin. Phys. C \textbf{46}, no.12, 123111 (2022).
		
		\bibitem{Guo:2023fih}
		F.~Guo and H.~S.~Li,
		``Analysis of the hidden-charm pentaquark states $P^{N^{0}}_{\psi}$ based on magnetic moment and transition magnetic moment,''
		[arXiv:2304.10981 [hep-ph]].
		
		\bibitem{Wang:2016dzu}
		G.~J.~Wang, R.~Chen, L.~Ma, X.~Liu and S.~L.~Zhu,
		``Magnetic moments of the hidden-charm pentaquark states,''
		Phys. Rev. D \textbf{94} (2016) no.9, 094018.
		
	
		
		
		\bibitem{Ozdem:2021ugy}
		U.~Ozdem,
		``Magnetic dipole moments of the hidden-charm pentaquark states: $P_c(4440)$, $P_c(4457)$ and $P_{cs}(4459)$,''
		Eur. Phys. J. C \textbf{81} (2021) no.4, 277.
		
		\bibitem{Ortiz-Pacheco:2018ccl}
		E.~Ortiz-Pacheco, R.~Bijker and C.~Fern\'andez-Ram\'\i{}rez,
		``Hidden charm pentaquarks: mass spectrum, magnetic moments, and photocouplings,''
		J. Phys. G \textbf{46} (2019) no.6, 065104.	
		
		\bibitem{Li:2024wxr}
		H.~S.~Li, F.~Guo, Y.~D.~Lei and F.~Gao,
		``Magnetic moments and axial charges of the octet hidden-charm molecular pentaquark family,''
		Phys. Rev. D \textbf{109} (2024) no.9, 094027.
	
	\bibitem{Lei:2024geu}
	Y.~D.~Lei and H.~S.~Li,
	``Radiative decay and axial-vector decay behaviors of octet pentaquark states,''
	Phys. Rev. D \textbf{110} (2024) no.5, 056026.
	
		
	
	
		\bibitem{Karliner:2015ina}
		M.~Karliner and J.~L.~Rosner,
		``New Exotic Meson and Baryon Resonances from Doubly-Heavy Hadronic Molecules,''
		Phys. Rev. Lett. \textbf{115} (2015) no.12, 122001.
		
	
		
		\bibitem{Nakamura:2021dix}
		S.~X.~Nakamura, A.~Hosaka and Y.~Yamaguchi,
		``$P_c(4312)^+$ and $P_c(4337)^+$ as interfering $\Sigma_c \bar{D}$ and $\Lambda_c \bar{D}^{*}$ threshold cusps,''
		Phys. Rev. D \textbf{104}  (2021) no.9, L091503.
		
		\bibitem{Ling:2021lmq}
		X.~Z.~Ling, J.~X.~Lu, M.~Z.~Liu and L.~S.~Geng,
		``$P_c$(4457) $\to P_c$ (4312) $\pi/\gamma$ in the molecular picture,''
		Phys. Rev. D \textbf{104} (2021) no.7, 074022.
		
		
		\bibitem{Wang:2018gpl}
		G.~J.~Wang, L.~Meng, H.~S.~Li, Z.~W.~Liu and S.~L.~Zhu,
		``Magnetic moments of the spin-$\frac{1}{2}$ singly charmed baryons in chiral perturbation theory,''
		Phys. Rev. D \textbf{98}, no.5, 054026 (2018).
		
		
		\bibitem{Coleman:1961jn}
		S.~R.~Coleman and S.~L.~Glashow,
		``Electrodynamic properties of baryons in the unitary symmetry scheme,''
		Phys. Rev. Lett. \textbf{6}, 423 (1961).


	
		\bibitem{Cheng:1997kr}
		T.~P.~Cheng and L.~F.~Li,
		``Why naive quark model can yield a good account of the baryon magnetic moments,''
		Phys. Rev. Lett. \textbf{80}, 2789-2792 (1998).
		
		\bibitem{Dahiya:2002fp}
		H.~Dahiya and M.~Gupta,
		``Octet magnetic moments and the Coleman-Glashow sum rule violation in the chiral quark model,''
		Phys. Rev. D \textbf{66}, 051501 (2002).
	
		
\bibitem{Li:2024jlq}
H.~S.~Li,
``Molecular pentaquark magnetic moments in heavy pentaquark chiral perturbation theory,''
Phys. Rev. D \textbf{109} (2024) no.11, 114039.
	
	

		
	\end{thebibliography}
\end{document}